# Room temperature triplet state spectroscopy of organic semiconductors

Sebastian Reineke[1,*] and Marc A. Baldo[1].

[1] Energy Frontier Research Center of Excitonics, Massachusetts Institute of Technology, 77 Massachusetts Avenue, Cambridge, MA 02139, USA.

[*] Author to whom correspondence should be addressed. Electronic mail: reineke@mit.edu.

## Abstract:

Organic light emitting devices and solar cells are machines that create, manipulate and destroy excited states in organic semiconductors. It is crucial to characterize these excited states, or excitons, to optimize device performance in applications like displays and solar energy harvesting. This is complicated if the excited state is a triplet because the electronic transition is 'dark' with a vanishing oscillator strength. As a consequence, triplet state spectroscopy must usually be performed at cryogenic temperatures to reduce competition from non-radiative rates. Here, we control non-radiative rates by engineering a solid-state host matrix containing the target molecule, allowing the observation of phosphorescence at room temperature and alleviating constraints of cryogenic experiments. We test these techniques on a wide range of materials with functionalities spanning multi-exciton generation (singlet exciton fission), organic light emitting device host materials, and thermally activated delayed fluorescence type emitters. Control of non-radiative modes in the matrix surrounding a target molecule may also have broader applications in light emitting and photovoltaic devices.

**Main Text:**

In contrast to most inorganic semiconductors, the excited states of organic molecules are highly localized, with the electron and hole densities typically confined to the same molecule. This localization and the consequently large exchange splitting generate two distinct states, referred to as singlets ($S_1$ – spin 0) and triplets ($T_1$ – spin 1). The energetic difference between $S_1$ and $T_1$ in organic molecules, the so-called singlet-triplet splitting $\Delta E_{ST}$, ranges from 50 meV to >1 eV (Refs. 1,2). Only the transition from the singlet state is spin-allowed, resulting in fluorescence. The triplet transition is considered a 'dark' state because it is quantum mechanically forbidden in the absence of perturbations such as spin-orbit coupling or vibrational coupling[3]. Unlike common inorganic semiconductors, where thermal energy is sufficient to mix between spin states efficiently, the energetics of the triplet states of organic materials play a determining role and must be considered with care. Organic light-emitting diodes are a prime example of the importance of the triplet state, because exciton formation through electrical injection is dictated by the spin statistics found in organic disordered films: for every singlet created, there are roughly three triplets formed[4]. This problem culminated in the development of heavy-metal complexes that allowed efficient phosphorescence from the triplet state through strong spin-orbit coupling[4]. To prevent quenching of the relatively long-lived triplet emission, it is necessary to adjust the triplet energy levels of all other organic materials within used in a phosphorescent-emitting organic device[5,6]. Other applications are similarly affected by the distinct, lower energy triplet states of the organic semiconductors used. In organic lasers, the triplet population has been proposed to be the principal impediment to the development of continuous wave lasing[7] and electrically pumped organic lasers[8]. Triplet states can act as a loss channel in solar cells[9,10], or as a pathway to longer-range exciton diffusion[11]. Recent developments show that novel phenomena like thermally activated delayed fluorescence[2] and multi exciton generation via singlet fission[12,13] can be realized through material design and especially the control of singlet and triplet energies. Thus, all major optoelectronic applications of organic semiconductors require spectroscopic characterization of the triplet states.

In a zero-order approximation, the transition from the excited triplet state $T_1$ to the ground state $S_0$ is disallowed, but – within a first-order approximation – gains finite probability taking vibronic and spin-orbit coupling into account[3]. Still, the radiative rate of phosphorescence $k_{r,P} \sim 10^{-1} - 10^2$ s$^{-1}$ of purely organic materials is very slow[3]. In comparison, the spin allowed transitions from $S_1$ to $S_0$ (fluorescence) are accompanied by orders of magnitude faster radiative rates of $k_{r,F} \sim 10^6 - 10^9$ s$^{-1}$. The same difference in transition probability is reflected in the absorption of light. Consequently, photo-excitation creates a vast majority of singlets, whereas direct triplet formation through absorption can be neglected. Fig. 1 shows an energy diagram of a typical organic molecule. Singlet and triplet states are split by the exchange energy $\Delta E_{ST}$. Singlet excitons generally find three channels: fluorescence, non-radiative deactivation to the ground state ($k_{nr,F}$) or the transfer to the triplet state via intersystem crossing (ISC). The latter is only possible, again, through vibronic and spin-orbit coupling[3]. Thus, the first requirement for observing phosphorescence is (i): $k_{ISC} > 0$. While many organic molecules fulfill this criterion of having finite ISC rates, some highly fluorescent dyes, e.g. laser dyes[14], exhibit negligible coupling outside the singlet manifold such that the triplet states cannot directly be photo-excited. Finally, the triplet deactivation is the competition between radiative ($k_{r,P}$) and non-radiative ($k_{nr,P}$) transitions in the triplet manifold (cf. Fig. 1). To observe phosphorescence, a second criterion needs to be met (ii): $k_{nr,P} \lesssim k_{r,P}$ The non-radiative losses can be divided into intramolecular and external losses to the environment[15], the latter being a result of coupling to phonon modes of the environment (e.g. host material) or collisions with quenching sites (e.g. oxygen or water). At room temperature, even in absence of oxygen and water, the non-radiative losses typically outcompete the radiative transition with rates of $k_{nr,P} \sim 10^2 - 10^6$ s$^{-1}$ (Ref. 3). Thus, phosphorescence at room temperature is only rarely observed and is often treated as a lab curiosity. To date, spectroscopy at cryogenic temperatures (77 K [liquid nitrogen], often even < 10 K) is the experiment of choice, simply because most, but not all non-radiative loss pathways (phonon-assisted) are 'frozen-out' so that criterion (ii) is met[3,6,16,17]. While triplet spectroscopy at cryogenic temperatures is a widely practiced, it remains a technique with many limitations. Cryo-cooling is also time-consuming and resource intensive, inhibiting rapid material screening. In addition, working with cryostats introduces physical constraints to the experiment, limiting the design freedom of the experimental setup and, sometimes introducing signal losses. Some key materials, like tetracene, undergo structural transformations below room temperature[18]. Other experiments to determine the triplet excited state energy are direct singlet ($S_0$) to triplet ($T_1$) absorption[19], indirect triplet depopulating quenching studies incorporating various acceptors of known triplet energy[20], and electron energy loss spectroscopy[21] – none of which have a high degree of accessibility and generality.

In this article, we report on the control of non-radiative modes within organic semiconductors that allows triplet state spectroscopy at room temperature for a wide variety of molecules. Using a polymeric, energetically inert host that is known to be best suitable for reducing non-radiative coupling losses to the environment[22], we find a surprisingly strong increase in the phosphorescence as a result of sample fabrication conditions. For a model compound [(BzP)BP] with noticeable phosphorescence at room temperature[23], we even observe – in contrast to the general trend – an increase of phosphorescence going from 77 K to 293 K. In addition, we discuss easily combinable concepts that increase the triplet population upon photo-excitation either through an indirect triplet sensitization using a triplet donor or through enhanced ISC through externally induced spin-orbit coupling.

## Results

**Highly rigid polymer films.** It is known that polymers, compared to other solid-state matrix materials (small molecules), have a high degree of rigidity, which makes them the material of choice to observe phosphorescence from organic lumophores[22]. Well-known candidates are PMMA and PS, which have triplet energies of 3.1 (Ref. 24) and 3.2 eV (Ref. 25), respectively, and are ideal wide-band gap host materials. They do not interfere with triplet states of visible or infrared emitting molecules. The rigidity of the final sample in which the organic materials are dispersed – typically in low concentration to avoid aggregation induced quenching[26] – is essential to suppress non-radiative deactivation of the triplet states through coupling to both intramolecular vibrational and rotational modes as well as to coupling to phonon modes of the embedding matrix. Thus, we explore ways to increase the system's rigidity by means of sample fabrication. The mixed film composed of the polymer host and small molecule organic guests is prepared by wet processing techniques (mostly by spin coating and in special cases by drop casting). A simple way to alter the formation of the thin films is the use of organic solvents with different vapor pressures/boiling points (b.p.). Here, the low vapor pressure solvent will keep the constituents dissolved longer, which allows a less disruptive transition to the solid phase. We have chosen chloroform ([chl], 21.09 kPa at 20 °C, b.p. = 61°C; vendor information) and methoxybenzene ([mb], 0.56 kPa at 25°C, b.p. = 61°C; vendor information) as high and low vapor pressure solvents, respectively. As an exemplar organic material we used (BzP)PB (chemical structure shown in Fig. 2a), for which we could recently show that its triplet state emission is very efficient at room temperature[23]. Hence, its phosphorescence can be used as a sensor for the degree of non-radiative deactivation in our samples. Fig. 2a and 2b plots photoluminescence (PL) transients of (BzP)PB at 77 and 293 K, following a quasi continuous wave excitation. The dye is embedded into a PMMA matrix with 2% by weight and spun from chloroform (Fig. 2a) or methoxybenzene (Fig. 2b). At times $t < 0$, the sample emits PL under constant illumination, where the intensity is scaled to the photoluminescence quantum yield (PLQY) of the respective sample. The PLQY at room temperature increases from 39 to 74% comparing the samples fabricated from chloroform and methoxybenzene. The excitation is turned off at $t = 0$ and, with respect to the temporal detection window of 600 ms, the prompt fluorescence immediately decays. The remaining delayed emission has its origin in the triplet manifold. For the film prepared from chloroform, a noticeable decrease of both its absolute intensity and decay lifetime is observed when increasing the temperature from 77 to 293 K. This clearly indicates that temperature assisted deactivation of the triplet state is dominant at room temperature, greatly reducing the phosphorescence signal from the sample. This picture dramatically changes for the thin film prepared from methoxybenzene. Here, the phosphorescence from (BzP)PB – indicated by the dashed lines in Fig. 2b – even increases roughly by a factor of 2 going from 77 to 293 K, presumably because the new technique has allowed us to image previously hidden processes such as the thermal dependence of intersystem crossing.

The initial non-linear decay of (BzP)PB in Fig. 2b is attributed to thermally activated delayed fluorescence (TADF) (Ref. 23). It is unknown, why the TADF signature is not observed when the film is spun from chloroform (cf. Fig. 2a). However, we suspect that the different sample preparation also affects the energetic disorder within the guest molecules, giving rise of coexistence of phosphorescence and TADF (in case of methoxybenzene solvent). The increase in PLQY from 39 to 74% cannot solely be explained by the enhanced phosphorescence, suggesting that the fabrication method also impacts the efficiency of the fluorescence channel (cf. Fig. 1).

In order to link the changes in observed phosphorescence with the sample preparation conditions, we acquired optical profiles of the thin films and determined densities of comparable polymer-only films fabricated under the same conditions. The sample topographies of both samples are shown in Fig. 2c and 2d for a scan area of hundreds of micrometers in lateral dimensions. While the film spun from chloroform shows a high amplitude in surface height modulation (surface roughness $R_{rms}$ = 64.9 nm), the preparation from the high boiling point methoxybenzene solvent yields a very smooth film, having a more than 1 order of magnitude reduced surface roughness of $R_{rms}$ = 1.9 nm. Using a quartz crystal microbalance as substrate for representative films of just the polymer, we can determine the relative change in density between the two fabrication methods. We find that $\rho_{mb} = (1.17 \pm 0.04)\rho_{chl}$. With this optimized sample morphology – showing strongly reduced non-radiative losses for the triplet emission – there is no necessity for experiments at cryogenic temperatures, because the photo-physics of the triplet emission remains almost unchanged in the temperature range between 77 and 293 K. The consequence is that phosphorescence spectra can be easily acquired at room temperature with greatly reduced effort.

It is worth noting that the same qualitative effect, *i.e.* a strong increase of phosphorescence compared to the sample spun from chloroform, can also be achieved for (BzP)PB by mixing a small organic molecule (here benzophenone, about 25% by weight) to the host polymer and keeping the solvent (chloroform) unchanged (Supplementary Fig. S1). Here, the additive is helping to suppress non-radiative modes either by affecting the local packing of the film or by frustrating the coupling to non-radiative modes within the solid.

In order to test this method, we have chosen organic small molecules for which phosphorescence spectrum has been reported at low temperatures but – to date – not at room temperature: TCTA[27], CBP[6,27,28], and NPB[6]. Their chemical structures are shown in Fig. 3c. (BzP)PB data is included this discussion for completeness. The samples are fabricated similar to (BzP)PB above, *i.e.* 2% by weight dispersed into PMMA, where the film is prepared by spin coating from methoxybenzene. Figure 3a shows the fluorescence of these materials obtained at room temperature under ambient conditions such that triplet excitons are effectively quenched by oxygen[20] and only the singlet emission is detected. Using a simple time gated setup (Methods) that allows for filtering the prompt emission (fluorescence), the phosphorescence of these materials spectra – shown in Fig. 3b – were measured at room temperature in a nitrogen atmosphere. The phosphorescence spectrum of (BzP)PB was obtained after additional processing steps correcting for the TADF contribution (details see Ref. 23). The triplet energies, obtained from the peak of the 0-0 vibronic transition, are 2.79 eV (TCTA), 2.58 eV (CBP), 2.37 eV [(BzP)PB], and 2.32 eV (NPB), all agreeing well with low temperature literature values[6,27,28].

These optimized fabrication conditions, leading to strongly enhanced phosphorescence at room temperature, have been used for all prepared samples that are the basis for the following results.

**Increasing and varying the triplet population.** The optical spectroscopy of 'dark' triplet states is often prone to the complication that triplets are not generated effectively, thus yielding a low phosphorescence intensity. In many cases, referring to Fig. 1, the population of $T_1$ is weak, because the sum of $k_{r,F} + k_{nr,F}$ simply outcompetes $k_{ISC}$ such that in addition to the weak nature of phosphorescence, triplets are also outnumbered by singlets being directly deactivated. In some cases, $k_{ISC}$ is even close to zero for which it is impossible to obtain phosphorescence, despite the possibility of a noticeable phosphorescence rate $k_{r,P}$. In order to overcome this bottleneck, we use the well-known organic material, benzophenone (BP), which can function as a versatile triplet sensitizer[29]. BP has an ISC efficiency close to 100% at room temperature[3] and a triplet energy of 2.96 eV (Supplementary Fig. S2), making it suitable for most organic materials having triplet levels in the visible and near infrared part of the electromagnetic spectrum. Thus, incorporating BP as an additive to a host guest system, absorption of light by BP yields singlets that are immediately converted into BP triplets, which then can transfer energy directly to the triplet state of the material of interest (acceptor) via Dexter energy transfer[30]. Here, the only possible major loss channel is the direct Förster resonant energy transfer (FRET) from BP singlet to acceptor singlet. However, the fast $k_{ISC} \sim 10^{11}$ s$^{-1}$ (Ref. 3) of BP suggests that ISC can be competitive to FRET.

Again, we have chosen (BzP)PB as a reference material for its known luminescent properties, having both contributions of TADF and phosphorescence to the luminescence at room temperature[23]. Furthermore, the fact that (BzP)PB has a small singlet-triplet splitting masks the phosphorescence in standard, steady state PL experiments simply because the spectral range of interest is flooded by fluorescence. In our study, we use PMMA as a host material to which we add BP with a weight ratio of 3:1. It is not possible to increase the content of BP further or even use it as a bulk host material, because it shows a tendency to crystalize as a result of strong van der Waals coupling between intermolecular benzene rings, leading to de-wetting of the substrate (Supplementary Fig. S3). Fig. 4a plots absorption data of various thin films differently composed of PMMA, BP, and the acceptor material (BzP)PB. For excitation wavelengths >250 nm, the absorption of PMMA can be neglected. While the absorption of (BzP)PB extends to 450 nm, the absorption cut-off of BP is at about 365 nm. This allows to probe the acceptor luminescence individually (excitation >400 nm) and, further, to create triplets on BP and vary their fraction to the total population by changing the excitation wavelength. The results are shown in Fig. 4b for low (77 K) and Fig. 4d for room temperature. At low temperature, the spectrum without sensitization (405 nm excitation) already shows weak shoulders at ~525 and 560 nm. Decreasing the excitation wavelength drastically changes the spectral shape of the (BzP)PB, with those features growing in to resemble the phosphorescence spectrum of (BzP)PB. Thus, clearly, the additive BP alters the fraction of singlets versus triplets in the acceptor dye towards more triplets, intensifying the phosphorescence. We have performed the same experiment at room temperature with the results displayed in Fig. 4d. Here, the same qualitative observation can be made: a red-shifted feature with respect to the fluorescence grows in with decreasing excitation wavelength. Thus, the triplet sensitizer BP strongly increases the triplet population and consequently enhances the radiative intensity of the triplet state. However, the vibronic features are washed out at ambient temperatures and the feature seems blue-shifted compared with the results obtained at 77 K.

In order to understand the spectral changes (with respect to the 77 K spectra) shown in Fig. 4d, we used our simple time gated photoluminescence setup (Methods) to solely measure the spectral distribution of the delayed component. We refer to this method as temporal isolation. Fig. 4c plots the delayed emission of 2% by weight (BzP)PB in the composite matrix (PMMA:BP[3:1]) at room temperature, excited with a 365 nm light emitting diode (LED) (black solid line). This delayed emission is composed of phosphorescence and TADF-type delayed fluorescence. For comparison, the red dash-dotted line is the phosphorescence only (cf. Ref. 23).

Hence, at room temperature, the delayed emission from (BzP)PB is a coexistence of TADF and phosphorescence, while the low temperature triplet emission is dominated by phosphorescence. Matching the high-energy side of the spectra obtained for different excitation wavelength and then subtracting the smaller from the larger spectrum yields a spectral distribution that is proportional to the luminescence originating from the triplet state – a method we call spectral isolation. The result for spectra obtained for excitations at 300 and 360 nm is plotted as black dashed line in Fig. 4c. The agreement of temporal and spectral data shows that, with the use of a triplet sensitizer – here BP –, the signature of the triplet luminescence can be easily determined from a continuous wave experiment even when fluorescence and phosphorescence strongly overlap due to a small exchange splitting (here: 290 meV). The true phosphorescence spectrum (cf. Fig. 4c) can be either obtained from time gated PL introducing a delay between the end of excitation and start of detection to sort out the TADF contribution or through spectral processing (cf. Ref. 23).

**Enhancing $k_{ISC}$ and $k_{r,P}$.** Spin orbit couplings between singlet and triplet excited states as well as between triplet excited and singlet ground states are the main driving force for $k_{ISC}$ and $k_{r,P}$, respectively[3]. Adding heavy metal atoms to the molecular core can greatly enhance the spin orbit coupling (scaling with atomic number $Z^4$) within a molecule[3,4], however such alteration of the molecular structure cannot be justified in the scope of determining triplet level of a specific molecule. The same qualitative effect can be achieved externally by introducing heavy atoms to the environment – the so-called external heavy atom effect. It has been made use of mostly in solutions and frozen solutions to experimentally access the triplet state[31]. Here, the key factor for its effectiveness is the close proximity between the heavy atom(s) and the center of the electronic transition, such that the electron density overlaps with the electric field of the heavy atom's nucleus[31].

As a practical example suitable to our sample fabrication process, we compare two polymers, PS and 4BrPS, where the latter has a bromine ($Z_{Br} = 35$) substitution in the 4-position of the side-chain benzene ring (cf. molecular structures in Fig. 5). We have chosen NPB (cf. Fig. 3) as organic lumophore, because its phosphorescence is very intense and singlet and triplet energies are spectrally well separated ($\Delta E_{ST} = 560$ meV). Figure 5 plots the PL transients of two samples, 2% by weight of NPB either doped in PS or 4BrPS, measured at room temperature following a quasi-continuous wave excitation (365 nm). Both samples show monoexponential decay of the phosphorescence with respective lifetimes of $\tau_{PS} = 400$ ms and $\tau_{4BrPS} = 108$ ms. The shortening of the lifetime $\tau = (k_{nr,P} + k_{r,P})^{-1}$ agrees with the observed increase in phosphorescence (cf. Fig. 5) relative to the fluorescence, suggesting an increase of $k_{r,P}$ of NPB induced by the brominated polystyrene matrix (4BrPS). With discrepancies between the change of lifetime $\tau$ (3.7-fold reduction) and changes in phosphorescence intensity (1.55-fold increase), the data clearly suggests a non-zero $k_{nr,P}$. This rate however cannot be quantified[23], because changes in intersystem crossing $k_{ISC}$, likely induced by the external heavy atom effect, are inaccessible from the data available.

**Determining triplet states of polyacenes.** Materials with very large exchange splitting have attracted much attention in the last years as potential candidates for singlet fission. This process is a spin-allowed multi exciton generation process where an excited singlet state splits into two triplet states[12]. The energetic requirement for this process is $\Delta E_{ST} \sim E_{T1}$ (cf. Fig. 1). Thus, for the identification of potential new fission materials, it is important to know, where the respective singlet and triplet energy levels are. Quantum chemical calculations are routinely used to determine a molecule's energy levels, but it is often the remaining uncertainty – up to 200 meV – of the calculation methods[32,33] that asks for more precise determination. This is especially important when the complexity of the molecules increases. In particular, singlet fission is very sensitive to the energetic separation of $S_1$ and $T_1$, where a few hundred meV determine whether the process is energetically downhill, resonant, or even uphill – with great impact of the fission dynamics[12].

We test two archetype singlet fission materials, anthracene and tetracene, with our techniques to spectroscopically determine their triplet energy levels at room temperature. We use 4BrPS as host polymer for its capabilities, as shown above, to enhance phosphorescence. Both anthracene and tetracene are dispersed into 4BrPS at 2 wt%. To enhance signal, rather than spin coating a film, we drop casted a thick, mixed film on a substrate, sitting at an elevated temperature (80 °C). It is important to note that such thick films will introduce significant re-absorption of the fluorescence of both anthracene and tetracene, owing to their small Stokes-shift. Thus, the high-energy side of the fluorescence spectra is artificially reduced in our measurements.

Figure 6a shows the PL spectrum obtained for anthracene at room temperature under continuous wave excitation 365 nm. Only a long pass, dielectric filter was used to measure the phosphorescence, sorting out higher order features from excitation and anthracene fluorescence. The 0-0 transition of the phosphorescence with peak energy of 1.83 eV agrees well with earlier reports[17,34]. For the phosphorescence, we find a triplet lifetime of 8 ms (cf. inset Fig. 6a).

For tetracene, even at 2 wt% concentration 4BrPS, we observe very intense delayed fluorescence (DF) (cf. inset Fig. 6b) with an very long time constant of 63 ms. Here, the delayed emission is not due to TADF but rather a result of triplet-triplet annihilation[35]. This indicates that (i) triplets are created in large densities through singlet fission as well as intersystem crossing, and (ii) the delayed emission is dominated by the spectral signature of tetracene's fluorescence. Furthermore, the fluorescence of tetracene is not the monomer but rather the aggregate emission[36]. We believe that the observations of delayed fluorescence and aggregate fluorescence suggest strong aggregation or even microcrystal formation of tetracene within the film. Such complex morphology has been observed before for tetracene embedded in a PMMA polymer[37]. In order to reduce the spectral contribution of the fluorescence to greatest extent, we employed a chopper wheel to filter out the prompt fluorescence (Methods). The PL spectrum of the delayed emission is shown in Fig. 6b. At 925 nm, corresponding to an energy of 1.34 eV, we observe a weak shoulder that we ascribe to the tetracene phosphorescence. Control experiments, not employing the chopper wheel, thus additionally measuring the prompt fluorescence, gave a decrease of this feature relative to the intensity of the fluorescence. The value of 1.34 eV agrees well with the triplet energy determined for tetracene in frozen solution (1.35 eV) at 77 K (Ref. 16). Comparison with the value obtained for a tetracene single crystal (1.25 eV, Ref. 19) suggests that the origin of the triplet emission of our sample must be tetracene molecules in dilute form. This result, as mentioned above, can only be understood assuming a complex nano- and microstructure of the sample, possibly containing isolated molecules as well as crystal domains. To our best knowledge, this is the first observation of tetracene phosphorescence at room temperature.

**Discussion**

Unlocking room temperature phosphorescence in the spectroscopy of triplet states and their dynamics allows for a significant simplification in experimental design and execution. The absence of cooling stages can yield more rapid measurements, such that multiple samples can be characterized in a manner of minutes. A reduction in non-radiative rates reduces the required excitation power per area, which noticeably reduces the risk of photo-bleaching the sample. Simple optical setups are sufficient. In our case, an optical fiber connected to a CCD spectrometer, or a photo-detector is placed in close proximity to the substrate edge, where the waveguided light is collected. This configuration makes use of geometrical gain $G$ (Ref. 38), as the sample functions similarly to a luminescent solar concentrator. At the same time, it makes complex focusing optics obsolete. The matching of refractive indices ($n$) between the glass substrate and the polymers used (all $n$ = 1.50±0.05) allows very efficient coupling between organic layers and glass substrate, such that the substrate edge homogeneously images the PL of the thin film. In our experiments, the geometrical gain is roughly $G$ = 6. Optical elements (chopper, filter, etc.) can be integrated in the sample-detector path without severe losses of signal intensity.

Our results show that readily detectable phosphorescence can be unlocked from the organic material's 'dark' state, if the dense packing induced by the sample fabrication affects the dominant loss modes of the molecule's triplet states. On the contrary, if the molecule of interest couples to modes, e.g. rotary motions of ligand arms[39] or intramolecular breathing modes[40] that cannot be frustrated in a rigid matrix environment, the observation of phosphorescence at room temperature remains a challenge. However, the methods to increase the triplet population through sensitization and speeding up the radiative rate from the triplet state through external heavy atom effects can equally well be applied at room temperature or low temperatures[3] (cf. Fig. 4b).

Future work should investigate in detail to what extent the formation of very rigid films compares to the effects induced by freezing the molecules to cryogenic temperatures. Further studies are also necessary to fully understand the structure-function relationship of the fabrication-induced morphology changes, especially to deduce pathways to even more effective suppression of non-radiative losses. The results presented here may also prove beneficial to the wider field of luminescent organic molecules, as emitting species always are in competition with non-radiative deactivation channels. In addition, the observation of simultaneous fluorescence and phosphorescence at room temperature enables new possibilities to study dynamic processes of organic semiconductors within the exciton's lifetime, which have been obscured to date by dominant non-radiative losses in the triplet manifold.

## Methods

**Sample preparation and characterization.** The materials used have acronyms as follows. PS: polystyrene, 4BrPS: Poly(4-bromostyrene), PMMA: Poly(methyl 2-methylpropenoate), BP: Benzophenone, CBP: [4,4'-bis(carbazol-9-yl)biphenyl], TCTA: [Tris(4-carbazoyl-9-ylphenyl)amine], NPB: [N,N'-di(naphtha-1-yl)-N,N'-diphenyl-benzidine], (BzP)PB: [N,N'-bis(4-benzoyl-phenyl)-N,N'-diphenyl-benzidine]. All materials, except tetracene, were used as received without further purification. Tetracene was purified by train sublimation. Vendor information: PS, 4BrPS, BP, tetracene, chloroform, methoxybenzene: Sigma Aldrich; CBP, TCTA, NPB: Lumtec Inc.; PMMA: Alfa Aesar, Anthracene: Fluka Analytical. Quartz substrates (1" by 1" by 1mm) were cleaned subsequently through a 5 min ultrasonic bath in acetone and a 2 min immersion (boiling) in isopropyl alcohol. If not stated otherwise, samples were either spun (spin speed of 2000 rpm (1000 rpm/s ramp) for 60 s) or drop-cast at elevated temperature (80 °C) from methoxybenzene solution in a protected nitrogen atmosphere (~2ppm oxygen and <2ppm water content). Relative densities of PMMA, spun from different solvents, were determined from sample thickness (stylus profilometer P-16+, KLA Tencor) and mass increase deposited onto a quartz microbalance using the Sauerbrey equation[41]. Surface topography images were taken with an optical profilometer (Wyko NT9100, Veeco).

**Optical spectroscopy.** Room temperature photoluminescence (PL) experiments was carried out in a nitrogen glovebox with ~2ppm oxygen and <2ppm water content, whereas measurements at 77 K were carried out in a bath of liquid nitrogen. PL spectra were recorded with fiber-coupled CCD spectrometers (visible: USB2000, OceanOptics; near infrared: SpectraPro 300i, Acton Research). Absorption data was recorded with a dual-beam spectrophotometer (Cary 500i, Varian). Excitation sources were either a 365 nm LED (Thorlabs) spectrally narrowed using a 365 nm, 10 nm FHWM band pass filter (Newport) or, for variable excitation wavelength, the monochromatic output of a high-intensity Xe-lamp (KiloArc, Optical Building Blocks Corp.). PL quantum yield measurements were done in a 6" integrating sphere (Labsphere) using a 405 nm laser (CPS405, Thorlabs) as excitation source and the radiometrically-calibrated USB2000 spectrometer as detector, following known protocols[42]. PL transients were recorded with an amplified silicon photo-detector with adjustable gain to match the long living PL of the samples (PDA36A, Thorlabs) attached to a fast oscilloscope (TDS 3054C, Tektronix) for read-out. The excitation source, a 365 nm emitting LED (Thorlabs), was driven by a pulse generator (B114A, Hewlett Packard), which also served as trigger source to synchronize the pulse train. Delayed PL spectra were collected similarly, using the USB2000 spectrometer as detector in a time-gated mode. In case of tetracene, a phosphorescope was build that allowed sample excitation while the detection path was blocked that allowed sorting out the prompt fluorescence using an optical chopper wheel (SRS540, Stanford Research Systems), running at 10 Hz, defining approximately 50 ms optical gates. To increase signal, the PL was detected from the edge of the glass substrate, making use of geometrical gain $G$ (here: $G \approx 6$) (Ref. 38).

photoluminescence quantum efficiency in organic solid-state thin films. *Jap. J. Appl. Phys.* **43**, 7729-7730 (2004).

## Acknowledgements

This work was supported as part of the Center for Excitonics, an Energy Frontier Research Center funded by the U.S. Department of Energy, Office of Science, Office of Basic Energy Sciences under Award Number DE-SC0001088 (MIT). SR gratefully acknowledges the support of the Deutsche Forschungsgemeinschaft through a research fellowship (Grant No. RE3198/1-1). The authors would like to thank Matthias E. Bahlke for the optical profilometer data and Trisha L. Andrew for advice on polymer processing.

## Author Contribution Statement

S.R. conceived the project and carried out all experiments and data analysis. M.A.B. supervised the project. Both authors discussed the results and jointly wrote the manuscript.

## Additional Information

The authors declare no competing financial interests.

**Figure Legends**

**Figure 1 | Competing rates of deactivation of a photo-excited organic molecule.** Lowest excited states in both singlet ($S_1$) and triplet ($T_1$) manifold are shown, which are energetically split by their exchange splitting $\Delta E_{ST}$. Singlets are created upon photo-excitation from the ground state ($S_0$) and can deactivate through fluorescence [F] and non-radiative recombination [nr,F] or transfer to the triplet state via intersystem crossing [ISC]. Competing rates for the triplet excitons are phosphorescence [P] and non-radiative recombination [nr,P]. Additionally displayed are the two criteria (i) and (ii) to obtain phosphorescence.

**Figure 2 | Rigidifying the solid-state sample through solvent selection.** Photoluminescence (PL) transients following a quasi-cw pump pulse indicating prompt (fluorescence) and delayed (thermally-activated delayed fluorescence and phosphorescence – see text for details) emission from (BzP)PB at 77 and 293 K. Date is scaled to the PL quantum yield of the respective samples under cw-illumination (red y-axis). Samples prepared from **a**, chloroform (b.p. = 61°C) and **b**, methoxybenzene (b.p. = 154 °C). Dashed lines in **b** indicate the phosphorescence contribution. **c**,**d**, Surface topography of thin films of just the host polymer (PMMA), spun from either chloroform [chl] (**c**) or methoxybenzene [mb] (**d**).

**Figure 3 | Singlet and triplet emission of organic semiconductors. a**, Fluorescence and **b**, phosphorescence from archetypical organic small molecules, chemical structures shown in **c**. All data obtained at room temperature.

**Figure 4 | Optical population of the triplet state. a**, Absorption of the various components of the [PMMA:BP]$_{3:1}$(BzP)PB 2 wt% sample. **b,d**, PL from a [PMMA:BP]$_{3:1}$(BzP)PB 2 wt% sample as a function of excitation wavelength, obtained at **b** 77 K and **d** 293 K. **c**, Triplet-originated emission at room temperature of [PMMA:BP]$_{3:1}$(BzP)PB 2 wt%, obtained by temporal or spectral isolation (see text for details). In addition, the (BzP)PB phosphorescence is plotted.

**Figure 5 | External heavy atom effect.** PL transients following a quasi-cw pump pulse indicating prompt (fluorescence) and delayed (phosphorescence) emission of NPB at room temperature using different host polymers, i.e. PS and 4BrPS. Inset shows the corresponding cw-PL spectrum.

**Figure 6 | Polyacenes – singlet fission materials. a**, Fluorescence and phosphorescence of anthracene obtained at room temperature under cw-illumination. The phosphorescence is recorded with additional 650 nm long pass filter to reject the fluorescence. Inset shows the phosphorescence decay with a lifetime of approximately 8 ms. **b**, Delayed emission of tetracene composed of delayed fluorescence (through triplet fusion) and weak phosphorescence. Prompt fluorescence is blocked using an optical chopper running with a frequency of 10 Hz. Inset shows the PL response upon 40 ms long 365 nm excitation pulse.

# Figures

## Figure 1

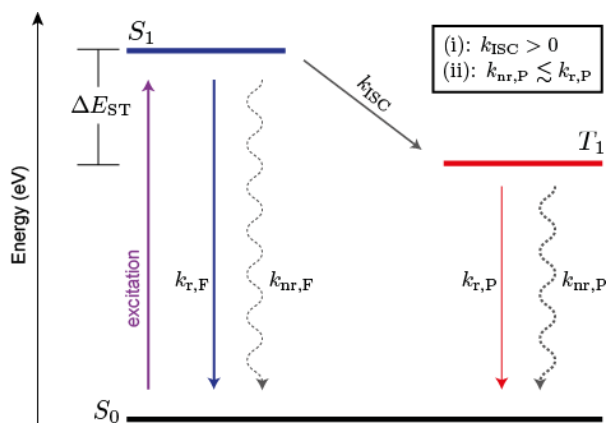

## Figure 2

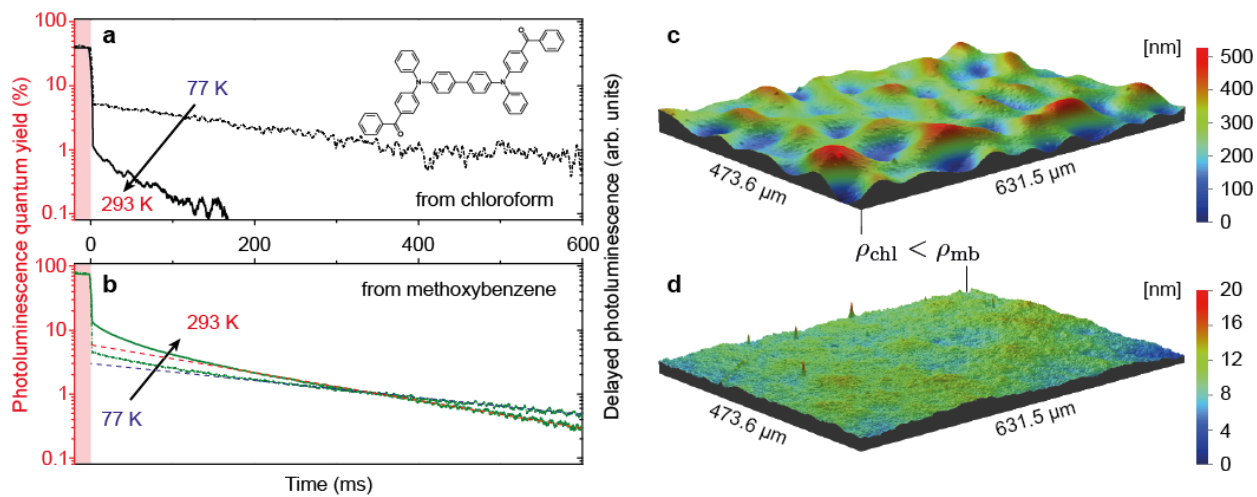

**Figure 3**

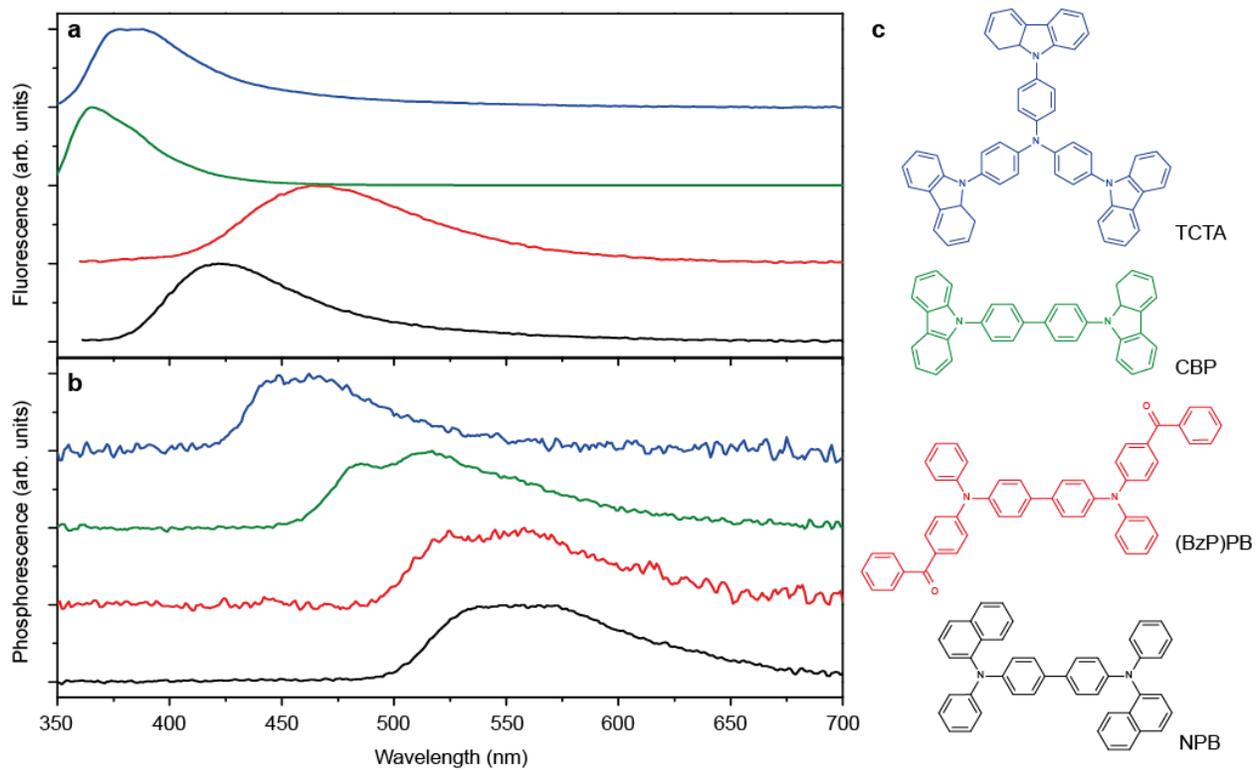

**Figure 4**

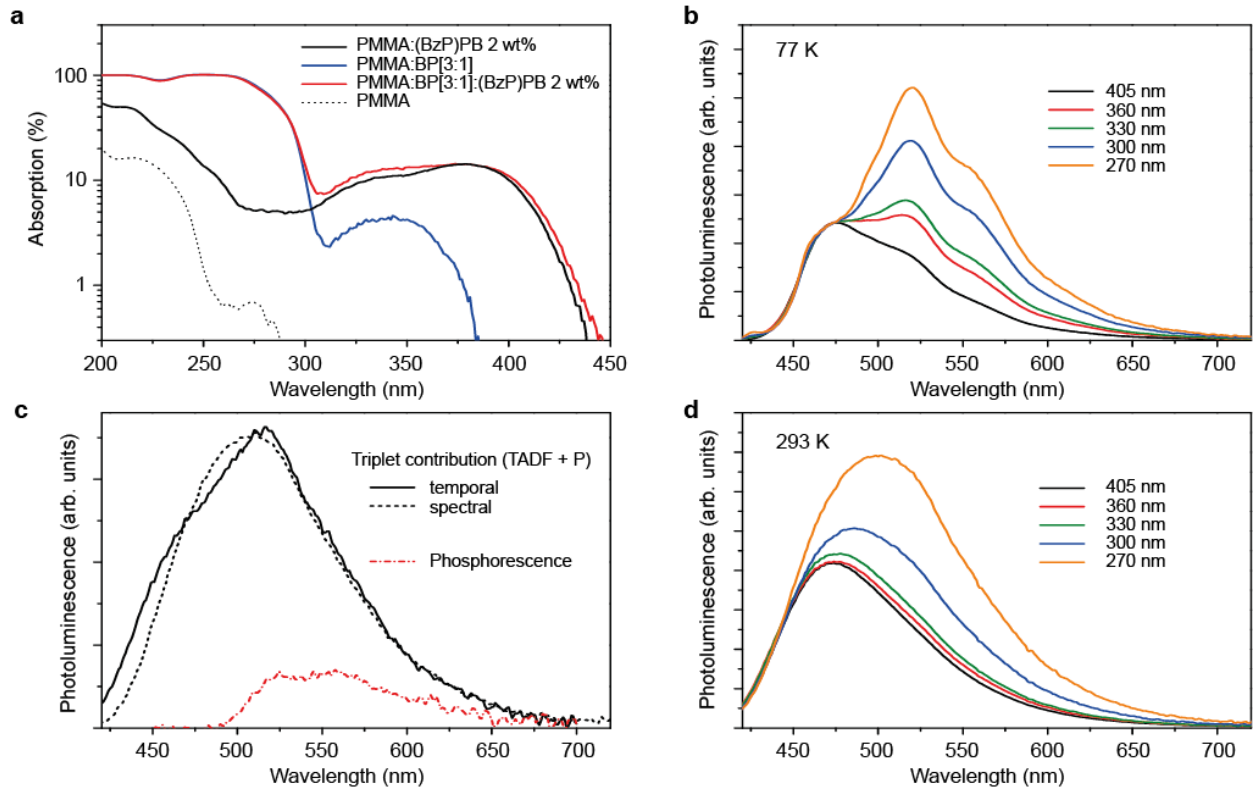

**Figure 5**

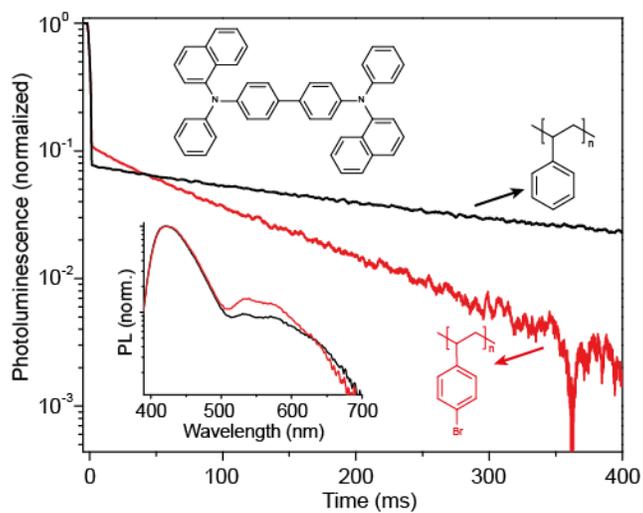

**Figure 6**

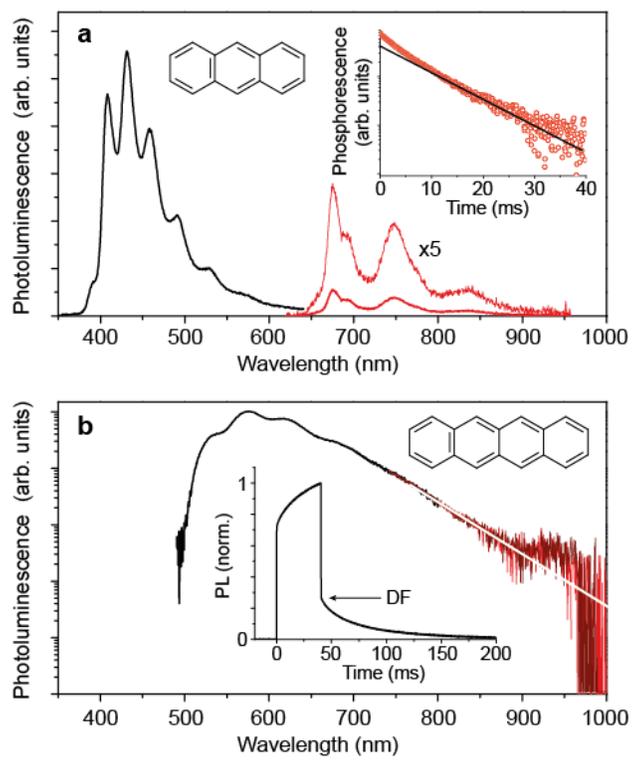